\documentclass{aa}
\usepackage{graphicx}
\newcommand\E[1]{\times10^{#1}}

\newcommand\U[1]{{\,\rm #1}}

\newcommand\al{\alpha}
\newcommand\gm{\gamma}
\newcommand\kp{\kappa}
\newcommand\lm{\lambda}
\newcommand\tht{\theta}
\newcommand\sg{\sigma}

\newcommand\rs[1]{_\mathrm{#1}}

\begin{document}

\title{The Crab Nebula at 1.3 mm}
\subtitle{evidence for a new synchrotron component\thanks{
Based on observations carried out with the IRAM 30-m telescope located at Pico
Veleta.
IRAM is supported by INSU/CNRS (France), MPG (Germany) and IGN (Spain).
}}

\author{
R. Bandiera\inst{1},
R. Neri\inst{2}
\and
R. Cesaroni\inst{3}
}

\offprints{R. Bandiera}

\institute{
Osservatorio Astrofisico di Arcetri, Largo E.Fermi, 5, 50125 Firenze, Italy\\
\email{bandiera@arcetri.astro.it}
\and
IRAM, 300 rue de la Piscine, 38406 St Martin d'H\`eres, France\\
\email{neri@iram.fr}
\and
Osservatorio Astrofisico di Arcetri, Largo E.Fermi, 5, 50125 Firenze, Italy\\
\email{cesa@arcetri.astro.it}
}

\date{Received ------ --, 2001; accepted ----- --, 2001}

\authorrunning{Bandiera et al.}
\titlerunning{The Crab Nebula at 1.3 mm}

\abstract{
We present the results of 1.3~mm observations of the Crab Nebula, performed
with the MPIfR bolometer arrays at the IRAM 30-m telescope.
The maps obtained, of unprecedented quality at these wavelengths, allow a
direct comparison with high-resolution radio maps.
Although the spatial structure of the Crab Nebula does not change much from
radio to millimetre wavelengths, we have detected significant spatial
variations of the spectral index between 20~cm and 1.3~mm.
The main effect is a spectral flattening in the inner region, which can be
hardly explained just in terms of the evolution of a single population of
synchrotron emitting electrons.
We propose instead that this is the result of the emergence of a second
synchrotron component, that we have tried to extract from the data.
Shape and size of this component resemble those of the Crab Nebula in X rays.
However, while the more compact structure of the Crab Nebula in X rays is
commonly regarded as an effect of synchrotron downgrading, it cannot be
explained why a similar structure is present also at millimetre wavelengths,
where the electron lifetimes far exceed the nebular age.
Our data, combined with published upper limits on spatial variations of the
radio spectral index, also imply a low-energy cutoff for the distribution of
electrons responsible for this additional synchrotron component.
Although no model has been developed so far to explain the details of this
component, one may verify that the total number of the electrons responsible
for it is in agreement with what predicted by the classical pulsar-wind models,
which otherwise are known to fail in accounting for the number of radio
emitting electrons.
This numerical coincidence can give indications about the origin of this
component.
We have also detected a spectral steepening at millimetre wavelengths in some
elongated regions, whose positions match those of radio synchrotron filaments.
The steepening is taken as the indication that magnetic fields in synchrotron
filaments are stronger than the average nebular field.
}

\maketitle

\section{Introduction}

The Crab Nebula is the prototype of synchrotron nebulae powered by a
spinning-down pulsar, also known under the name of ``plerions'' (Weiler \&
Panagia \cite{wei78}).
This is an extensively studied object, and a wealth of information on the
synchrotron nebula comes from detailed observations performed in various
spectral ranges, like in radio, infrared, optical, UV and X rays.

Modelling all the available data in a comprehensive frame represents a
formidable task for the theory.
Classical approaches to the modelling of the Crab synchrotron emission, like
Pacini \& Salvati (\cite{ps73}) and Kennel \& Coroniti (\cite{kc84a},
\cite{kc84b}), got some success.
But more elaborate models can hardly get any
substantial improvement with respect to the original approaches, partly because
the geometric structure of the Crab Nebula is very complex, but probably also
because the processes involved are not fully understood.
When more quantitative and detailed modelling will be possible with other
plerions we expect to face similar problems: in these respects a large part of
the results on the Crab Nebula are likely to be exported to other objects.

Considering just the total luminosity spectrum, Pacini \& Salvati
(\cite{ps73}) successfully reproduced it from radio to optical, by simply
assuming a pure power-law distribution for the injected ``particles''
(hereafter used to indicate relativistic electrons, as well as positrons): but
in order to explain by their model the further spectral steepening in the X-ray
range and beyond, an {\it ad hoc} spectrum for the injected particles is
required.
On the other hand Kennel \& Coroniti (\cite{kc84b}) successfully reproduced the
spectrum from optical to gamma rays, just assuming a power-law distribution
(over a range of energies) for the particles accelerated at the termination
shock of the pulsar relativistic wind.
However their model fails in explaining the observed radio emission: the
problem is that the best-fit wind model implies also an estimate of the total
number of radio emitting particles injected into the nebula, which is at least
a factor 100 lower than what measured.
Up to now this discrepancy has been cured only by introducing some {\it ad hoc}
assumptions (see e.g.\ Atoyan \cite{ato99}).

As far as the spatial structure of the nebula is concerned, it is not difficult
to explain qualitatively its behaviour with frequency, namely the shrinking of
the nebular size with increasing frequency: in fact the latter corresponds to
increasing particle energy, and therefore decreasing synchrotron lifetimes.
However quantitative approaches fail to reproduce the observed profiles, both
in the Kennel \& Coroniti (\cite{kc84b}) and in the Pacini \& Salvati
(\cite{ps73}) frameworks: the implications of the assumptions in the latter
paper on the nebular spatial extent have been investigated by Amato et al.\
(\cite{aea00}).
A common characteristic of the above models is that the particles are advected
outwards with the magnetic field, following the MHD equations.
Somehow better results are for instance obtained by including also diffusive
processes, but only when an {\it ad hoc} diffusion coefficient is taken.

An alternative to the above scenarios relies on assuming the coexistence of two
(or more) components of injected particles, with different spectra as well as
with different spatial locations.
But one may be unwilling to increase the complexity of the models, unless a
stringent evidence in that sense comes out from the observations.
Since adiabatic losses preserve the slope of the particles distribution, the
most direct test on the presence of multiple components of the injected
particles consists into measuring spatial variations of the synchrotron
spectral index that cannot in any way result from a synchrotron downgrading
(i.e.\ the spectral softening consequent to synchrotron evolution of the
emitting particles).
This can be done observing at frequencies so low that the related particles are
subject to negligible synchrotron losses, and therefore whose distribution
retains the slope which had at the injection.

Beforehand this kind of test had been done only at radio wavelengths (below
with the term ``radio wavelengths'' we roughly indicate wavelengths above
1~cm).
Previous claims of spatial variations of the radio spectral index (Velusamy et
al.\ \cite{vea92}) have been then contradicted (Bietenholz et al.\
\cite{bea97}).
Possibly some variations of the spectral index are present in the very central
region, with scales of a few arcsec and associated to the ``wisps'' structures
(Bietenholz \& Kronberg \cite{bk92}): however such result could be a mere
artifact, originated by comparing data taken at different epochs, in the
presence of rapidly moving wavy structures (Bietenholz et al.\ \cite{bea01}).
In fact recent radio observations (Bietenholz et al.\ \cite{bea97}) strongly
support the idea of a single injected distribution, by putting a tight upper
limit, 0.01, to spatial variations of the spectral index, at least on scales
larger than 16\arcsec.

In this paper we will show that millimetric wavelengths represent the most
appropriate spectral range to investigate this issue in the Crab Nebula, by
providing new pieces of information with respect to the radio.
The map presented in this paper, with a 10.5\arcsec\ resolution, is by far
better than the only map of the Crab Nebula previously published at these
wavelengths (Mezger et al.\ \cite{mea86}, with only 120\arcsec\ resolution).
The paper is organized as follows: in Sect.~2 we report on the observation
parameters and on the data reduction; Sect.~3 describes the procedure by which
our 1.3~mm map has been compared with a 20~cm radio map; features coming out
from this comparison, namely the emergence of a second component in the inner
regions and a general bending in filaments spectra, are respectively discussed
in Sect.~4 and Sect.~5; Sect.~6 shows that the nature of the emission from the
inner component is synchrotron; the morphology of the new synchrotron component
is compared in Sect.~7 with maps at other wavelengths; in Sect.~8 we comment on
possible spurious effects on our results deriving from time variability of the
source; Sect.~9 concludes.

\section{Observations}

The Crab Nebula has been observed during two different runs at the 30-m
telescope located at Pico Veleta (near Granada, Spain) and operated by the
Institute de Radioastronomie Millim\'etrique (IRAM); to this purpose, we have
used two different bolometer arrays (MAMBO) developed by the
Max-Planck-Institut f\"ur Radioastronomie (MPIfR).
In the former run (December 2--3 1998), characterised by poor atmospheric
conditions, we have mapped the source using the 19-element bolometer; while in
the latter (February 13--14 2000) we have obtained a map with the newly
developed 37-element bolometer and with good atmospheric conditions.

In both the 19-element and the 37-element array the individual bolometers are
organized in a hexagonal structure.
The FWHP beamwidth of the telescope at 1.3~mm has been estimated from pointing
scans on NRAO\,530 and found to be $\sim\!10\farcs5$.
The frequency bandwidth ranges from 200 to about 280~GHz:
the effective frequency of
the observation varies slightly with the atmospheric opacity and with the
spectral index of the source, and keeps typically in the range 230--240~GHz.
A detailed description of the instruments may be found in Kreysa et al.\
(\cite{kgg98}).

The maps have been obtained by on-the-fly scans in the dual-beam mode.
The telescope scans the source in azimuth (at the velocity of 4\arcsec/sec,
corresponding to a spatial sampling of 2\arcsec), while the secondary mirror is
chopping, again in azimuth, at a frequency of 2~Hz with an elongation of
45\arcsec.
The shift in elevation between subsequent elementary scans is 4\arcsec: in a
complete raster scan each element of the bolometer then samples the whole
source.
Various slices of the Crab Nebula (4 for the December 1998 run; 8 for the
February 2000 one) have been observed, each at a
slightly different parallactic angle and center. To provide the
best mapping coverage, the shape and size of the slices were
continuously optimized during the observations. Redundant
information was used to improve the signal-to-noise ratio, as well as to
keep trace of the quality of the observations. The slices were
gridded on a regular equatorial frame and finally combined
according to their weights. In the February 2000 run the mapping sensitivity
varies by a factor of two over the whole plerion: it is found to be
highest close to the center, in the north and southeast. Outside
the plerion, the rms noise sensitivity is about 6~mJy/beam.

The atmospheric zenith opacity has been monitored by repeating a skydip every
hour.
The flux calibration was performed on planets.
The pointing was checked every hour on QSO~B0528+134
and found to be better than
2\arcsec.
Given the superior quality of the data taken in February~2000 under good
atmospheric conditions, in the following we will use only the resulting
average map obtained with the 37-element bolometer array: this is shown in
Fig.~\ref{fmap}.

The map obtained in December 1988, though of poorer quality, has
been used to check the accuracy of the absolute calibration.
Though differences in the overall calibration have not shown up at
levels above 10\% between the December 1998 and February 2000 maps, we prefer
to be more conservative and claim rather a 20\% accuracy for the
map of the Crab Nebula. This value also includes
possible flux variations across the map (at a $\sim\!200\arcsec$ scale)
due to differential pickup emission between the on and off
error-beams of the telescope, but is, according to simulations, by
no means affecting quantitatively the relative accuracy at lower
mapping scales. The relative accuracy across the map has
independently been estimated also by comparison of two map slices taken
under different observing conditions in February 2000 and turns
out to be $\sim\!10$\%.

The data have been reduced using the NIC package, which is part of the GAG
software developed at IRAM and Observatoire de Grenoble.

\begin{figure}
\caption{
Map of the Crab Nebula at 1.3~mm with $10\farcs5$ resolution.
The contour levels range from 0.2 to 6.8 in steps of
0.6~$\U{mJy\,arcsec^{-2}}$.
}
\label{fmap}
\end{figure}

\section{Comparison with a 20~cm map}

In this section we describe the procedure we have followed in order to estimate
the difference between our 1.3~mm map and a reference radio map at lower
frequency.
Due to the high homogeneity of the spectral index over the radio nebula
(Bietenholz et al.\ \cite{bea97}) the choice of the reference radio frequency
is not crucial.
We have used a map of the radio continuum at 1.41~GHz, based on data taken at
the NRAO Very Large Array (VLA), and kindly provided by M.F.~Bietenholz.
This is one of the highest-quality maps of the Crab Nebula produced so far in
the radio range.

The VLA data, their analysis and the quality of the resulting map are described
in detail by Bietenholz \& Kronberg (\cite{bk91}).
In short, the data have been collected with a series of runs in 1985--1986.
The map has been produced by combining observations in all 4 VLA
configurations: the cleaned beam size of this map is $2\farcs0\times1\farcs8$
and the $u$-$v$ plane results very well sampled.
Therefore this map (which is also corrected for primary mean attenuation)
guarantees a high photometric accuracy also on large
scales.
The integrated flux in the cleaned map is 870~Jy, close to the expected
907~Jy:  we have then increased by 4.3\% the flux of this map, in order to
correct for the small flux deficiency produced by residual gaps in the $u$-$v$
coverage.

In order to allow an accurate comparison between maps taken at different
epochs, we have first corrected the size and flux for the secular evolution.
We have taken 1986 as the average date of the VLA map, and 2000 for our map:
this time difference results into an estimated size difference of 1.9\% (for an
expansion time of 730~yr, Bietenholz et al.\ \cite{bea91}) and a flux
difference of $-$2.3\% (for a fading rate of $-0.167$\%$\U{yr^{-1}}$,
Aller \& Reynolds \cite{ar85}).
The total flux of the resulting radio map, extrapolated to year 2000, is then
886~Jy.
We have finally convolved this image with a gaussian function in order to
obtain a $10\farcs5$ resolution map.

In the 1.3~mm map we have evaluated in $\simeq$5.5~mJy/beam
the average background
level in the region external to the area covered by the Crab Nebula (where the
boundaries of the nebula have been obtained from the smoothed radio map).
This offset accounts for a flux of $\simeq$4.4~Jy
on the area covered by the nebula,
and of $\simeq$21~Jy
over the total observed field: this is a small quantity, in the
sense that its presence does not change the qualitative scenario.
However, in order to obtain more accurate quantitative results, we have
subtracted this offset from the 1.3~mm map, by assuming that this background
term is purely additive.
After this correction the total flux of the Crab Nebula at 1.3~mm is estimated
as $\simeq$260~Jy
(still taking 20\% as a conservative estimate of the uncertainty in
the absolute flux).

After this processing, at a first glance the maps at 1.3~mm and 20~cm look
quite similar, but a quantitative comparison shows significant differences.
The most standard way to outline such differences is generating a map of the
spectral index (Fig.~\ref{spindx};
where the spectral index $\al$ is defined by
$F(\nu)\propto\nu^\al$).

\begin{figure}
\caption{
Map of the spectral index $\al$, for the 20~cm -- 1.3~mm range.
Contour levels range from $-$0.28 to $-$0.20
in steps of 0.02 (correspondingly
the grey scale ranges from black to white).
}
\label{spindx}
\end{figure}

The most apparent effect, in this map, is a flattening of the spectral index in
the inner regions.
The level of spectral index inhomogeities is one order of magnitude
larger than the upper limit put on the radio spectral index (Bietenholz et al.\
\cite{bea97}).
For reference, a change of 0.01 in the spectral index corresponds to a 5\%
change in the flux ratio between the two maps: therefore the absolute flux
correction we have applied on the VLA map could contribute at most for a 0.01
on large scales, while the (conservative)
20\% absolute photometric accuracy for the
MPIfR bolometer array could account at most for a 0.04 offset in the spectral
map.
The relative photometric accuracy on smaller scales is only about 10\%,
and therefore spectral index variations of $\sim\!0.05$ over
distances of about 1\arcmin, as seen in Fig.~\ref{spindx}, cannot be ascribed
to mere instrumental effects.
A similar, although somehow noisier spectral map is obtained using the
December 1998 data: therefore we exclude that these spectral index variations
are just an effect of peculiar observing conditions.

There are two more potential sources of bias, related to instrumental
characteristics.
Bietenholz et al.\ (\cite{bea97}) mention a correlator nonlinearity in the
radio map, that could in principle be responsible for artifacts: however they
also specify that this bias may affect only angular scales $>2\farcm7$, while
in our spectral map most features are at smaller scales; we have also verified
that the local measured spectral index may differ between regions with the same
surface brightness: this cannot be accounted for as a mere effect of
nonlinearity.
Moreover a gaussian convolution may not accurately reproduce the Point Spread
Function (PSF) relative to the IRAM map, and in the case the radio map has
been oversmoothed a slightly flatter spectrum is expected at the top of the
intensity peaks: however, even by increasing by factor 50\% the resolution of
the the radio map with the respect to the estimated value for PSF equalization,
we obtain a spectral map
which is still qualitatively consistent with what shown in Fig.~\ref{spindx}.
We therefore conclude that no substantial bias derives from any of the effects
discussed above.

In the following, while discussing the structure of the flatter-spectrum inner
region, as well as of some steeper-spectrum elongated structures, we shall
recognize the spatial coincidence with structures revealed at other
wavelengths: we take these as two ``a posteriori'' arguments still in favour of
the reality of the observed inhomogeneities in the spectral map.
A final concern refers to the possibility that the comparison of observations
taken at different epochs actually traces changes in time: anyway, as it will
be discussed in Sect.~8, not even that effect can account for the measured
spectral index inhomogeneities.

\section{Evidence for the presence of additional components in the 1.3~mm map}

The spectral map in Fig.~\ref{spindx}, considered together with the radio
spectral map (Bietenholz et al.\ \cite{bea97}), indicates that in some regions
of the nebula the average spectral index between 20~cm and 1.3~mm wavelengths
is flatter than that at 20~cm.
This means that in these regions the spectrum shows a positive
second derivative (in a log-log plot) in the wavelength range between radio and
millimetric (mm) wavelengths: i.e.\ $F(\nu)\propto\nu^{\al(\nu)}$
where $\al$ is increasing with $\nu$.
A similar spectrum is unusual for a single synchrotron component, unless the
injection spectrum itself has a positive second derivative.
The effect of synchrotron evolution is instead that of steepening the spectrum
a smaller wavelengths, thus producing a spectral behaviour opposite to what we
have measured.

From the spatially integrated spectrum of the Crab Nebula it is well known that
a spectral break occurs at wavelengths around $20\U{\mu m}$.
Such break is commonly accepted to be an evolutive one, so that at wavelengths
longer than the break the emission originates from particles whose evolution is
not appreciably affected by synchrotron losses.
In the case of adiabatic evolution, the particle distribution preserves its
slope.

Therefore the most natural conditions leading to flatter spectra at smaller
wavelengths in some regions is that, in that spectral range and in those
regions, also the emission from a further population of particles is detected.

We shall thus attempt to map this secondary component, hereafter referred to as
``component {\bf B}'', while we reserve the label ``{\bf A}'' for the component
dominating in flux at radio wavelengths.
With this in mind we have subtracted from our mm map an extrapolation of the
radio map down to 1.3~mm, by adopting a power-law index $-$0.27, as derived
from multiwavelengths radio measurements of the total flux (Kovalenko at al.\
\cite{kea94}).
The map of the residuals is presented in Fig.~\ref{resid}: the main feature in
this map is a component located in the central region of the Crab Nebula; a
further component consists of some elongated features with emission
depressed with respect to their surroundings. They are better visible on the
N side of the nebula, where even negative values are present in the map of the
residuals.

\begin{figure}
\caption{
Map of the residuals, after subtracting the extrapolation of the radio map with
index $-$0.27.
The contour levels range from $-$0.4 to 1.7 in steps of
0.3~$\U{mJy\,arcsec^{-2}}$.
Negative contours are indicated by dashed lines.
The thick lines follow qualitatively the locations of some radio
filaments (see Fig.~\ref{fila}),
which in this map correspond to local depressions.
}
\label{resid}
\end{figure}

At a first glance these features may be interpreted as just a noise effect.
But a more careful analysis shows that they are well correlated with the
positions of synchrotron filaments, visible in the radio image.
In order to settle more quantitatively this effect we have synthesized a map of
the filamentary component (hereafter labelled as ``{\bf f}'', see
Fig.~\ref{fila}), that together with the ``amorphous'' component (hereafter
labelled as ``{\bf a}'') is responsible for the structures visible in the radio
map (i.e.\ {\bf f}+{\bf a}={\bf A}).

The method as well as the parameters that we have used to synthesize {\bf f}
are described in Appendix A.
They have been tuned according to the following requirements: component {\bf a}
must look very smooth; component {\bf f} must instead contain mostly high
spatial frequencies; negative peaks must neither appear in the map of component
{\bf f} nor in that of {\bf a}; the total flux of the filaments component must
be as high as possible, within the limits set by the previous constraints.
Even with these requirements some arbitrariness is left on the choice of the
procedure, but the results are not strongly dependent on this choice.
The resulting map of filaments, convolved to the resolution of our 1.3~mm map,
is shown in Fig.~\ref{fila}.

\begin{figure}
\caption{
Map of synchrotron filaments (component ``f''),
extracted from the radio map with a procedure
described in Appendix A, and downgraded to $10\farcs5$ resolution.
The contour levels range from 1.5 to 10.5 in steps of
1.5~$\U{mJy\,arcsec^{-2}}$.
The positions of some filaments correlate with the negative features in the
residual map
(the thick white lines correspond to those drawn in
Fig.~\ref{resid}).
}
\label{fila}
\end{figure}

We have then estimated the relevance of the filamentary structure at 1.3~mm
wavelength, by deriving the best-fit linear combination of components {\bf f}
and {\bf a}, which minimizes the ripples in the residuals.
The factor found for {\bf a} has here little meaning, because
component {\bf a}, which accounts only for the large-scale structures, is
rather insensitive to the morphological differences between component {\bf A}
and component {\bf B}. Instead we shall concentrate on the best-fit factor
(0.15) we have found for {\bf f} (this component
accounts for the small scale fluctuations, and is highly sensitive to the
pattern to match): its value is only $\sim\!$60\% of what
expected for a pure power-law extrapolation from radio wavelengths, with a
$-$0.27 spectral index.

The physical meaning of this deficiency will be discussed in the next section.
Here we use this result just to obtain a ``cleaned image'' of component {\bf
B}.
The factor by which the map of component {\bf a} must be subtracted is however
rather uncertain.
As a first choice let us take an extrapolation from the radio with the
canonical index $-$0.27: the result is shown in Fig.~\ref{compb}.

\begin{figure}
\caption{
Map of ``component {\bf B}'', after subtracting the {\bf a} (amorphous)
component, extrapolated from the radio with index $-$0.27, and the {\bf f}
(filaments) component, using the best-fit factor 0.151
from the radio.
The contour levels range from 0.2 to 2.3 in steps of
0.3~$\U{mJy\,arcsec^{-2}}$.
}
\label{compb}
\end{figure}

According to this figure, component {\bf B} presents a brighter inner part,
which is dominated by a feature elongated by about $1\farcm6$ in the NE-SW
direction, and comprises two fainter extensions in the SE and NW sides, for an
overall size of $1\farcm6$ also in that direction.
This structure resembles that observed in X rays: this point will be discussed
in Sect.~6.
Beyond the inner part, Fig.~\ref{compb} also presents an outer halo, extending
over an area of about $4\arcmin\times5\arcmin$.

Uncertainties on the most appropriate factor in the subtraction of component
{\bf a} reflect into uncertainties on the structure and on the integrated flux
of the halo of component {\bf B}, while the structure of its inner part is
rather insensitive to the details of the subtraction procedure.
For this reason at present we can only give a coarse estimate of the total flux
of this component, because dominated by its halo: on the other hand the
structure of the inner part is well outlined by Fig.~\ref{compb}.

Using for component {\bf a} an extrapolation from radio with index $-$0.27,
the
estimated integrated flux of component {\bf B} results in 59~Jy;
on the other
hand $-$0.26
is the maximum index for which after the subtraction a positive
map
is left: this value is consistent both with the quoted uncertainty on the radio
spectral index (0.04; Kovalenko et al.\ \cite{kea94}) and with that on the
1.3~mm absolute photometry (20\%, equivalent to 0.04 on the spectral index).
With a $-$0.26 extrapolation the integrated flux of this component is measured
in 54~Jy.
The integrated flux of the inner part is anyway much lower: assuming that the
halo emission presents a plateau in the central region, we estimate a residual
flux of about 2.2~Jy.

\section{Spectral bending in the filaments spectra}

As mentioned in the previous section, at 1.3~mm the fluxes of the filaments are
on the average 0.60
times lower than the values extrapolated from the radio
data.

Our first concern has been to examine whether this is a reliable result, or if
it can be just the effect of biases in the components subtraction.
For instance, a spurious effect may take place if the PSF equalization between
the radio and mm maps has not been carried on properly: if we have
understimated the width of the 1.3~mm PSF, the filaments at 1.3~mm would look
less prominent (relative to the amorphous component) than in the corresponding
PSF-equalized radio map.

In any case such an instrumental bias cannot account for what is observed.
First of all, in the case of a bias of this kind, we expect to see in the map
of residuals (Fig.~\ref{resid}) negative peaks near the centre of the
filaments, and positive residuals on the filament sides, to equilibrate the
flux: but we did not measure any such effect.
Even using an artificially smoothed radio map (with a PSF 50\% larger than
required) we still came out with basically the same result as before.

Therefore the relative weakness of the filaments in the map at 1.3~mm is a
rather robust and safe result.
It is in agreement also with maps of the optical continuum, in which the
filaments structure is on the average much less pronounced than in the radio
(Wilson \cite{wil72b}).

We have found two different explanations for this effect, one based on the
energy evolution of particles confined in individual filaments, and the other
based on the relative efficiency of particle diffusion from filaments.

\subsection{Energy evolution of confined particles}

Since at radio wavelengths the spectral index of the filaments does not differ
from that of the amorphous component (Bietenholz et al.\ \cite{bea97}), the
weakening of filaments at mm wavelengths could be accounted for by assuming
that a spectral break intervenes in between radio and mm wavelengths.

In the case of a steady injection an evolutive break is produced at the
wavelengths at which, for the emitting particles, adiabatic and synchrotron
losses are comparable: in this case the expected change for the spectral
index of the energy distribution is 1 (see e.g.\ Pacini \& Salvati
\cite{ps73}).
Let us thus model the synchrotron emission from electrons with a bi-power law
energy distribution, with slopes $-$1.54 (in order to match the measured radio
spectral index) and $-$2.54 respectively.

This represents a simplified case; but the further smoothing, due to the fact
that the synchrotron spectrum for a single particle is broad band, causes the
derived radiation spectrum to be anyway not far from what obtained with a more
detailed modelling.
In this way we derive for the filaments a break frequency ($\nu\rs{b}$) of
$\sim$80~GHz,
namely a factor 200 lower than in the integrated Crab spectrum.

From this one may infer the typical magnitude of the magnetic field ($B$) in
the filaments.
If particles are confined into individual filaments the energy ($E\rs{b}$)
corresponding to the spectral break is proportional to $B^{-2}$; in turn
$\nu\rs{b}\propto BE\rs{b}^2\propto B^{-3}$: therefore filament magnetic fields
are typically $\sim\!200^{1/3}\sim\!6$ times higher than the average nebular
field.
The dynamical implications of this field excess in the filaments are beyond the
scope of the present work.

\subsection{Particle diffusion through filaments}

A very unlikely result follows if all particles are effectively diffusing
through filaments.
In this case $E\rs{b}$ would be the same as in the interfilament medium and
therefore $\nu\rs{b}\propto B$, leading to estimate a field in the filaments
much lower than in the rest of the nebula: such a scenario, among others, would
not account for the enhanced synchrotron emission in filaments, and we
therefore reject it.

However particle diffusion may give more relevant effects to this problem, if
the transition to the diffusive regime occurs in between radio and mm
wavelengths.
The diffusion time, $t\rs{D}=r\rs{F}^2/\kp$ can be evaluated as:
\begin{equation}
  t\rs{D}\simeq19,000\U{yr}
    \left(\frac{\tht\rs{F}}{1\arcsec}\right)^2\!\!
    \left(\frac{B}{0.3\U{mG}}\right)^{3/2}\!\!
    \left(\frac{\lm}{1.3\U{mm}}\right)^{1/2}\!\!
    \left(\frac{\kp}{\kp\rs{B}}\right)^{-1}\!\!\!\!\!\!\!\!,
\end{equation}
where $r\rs{F}$ and $\tht\rs{F}$ are respectively linear and angular transverse
size of a filament
(we have used 2~kpc for the distance of the Crab Nebula),
$\kp\rs{B}\simeq(m\rs{e}c^3\gm)/(3eB)$ (where $m\rs{e}$ and
$e$ are respectively mass and charge of the electron, $c$ is the speed of light
and $\gm$ is the particle Lorentz factor) is the Bohm diffusion coefficient,
which evaluates:
\begin{equation}
  \kp\rs{B}=1.5\E{21}\U{cm^2s^{-1}}
    \left(\frac{B}{0.3\U{mG}}\right)^{-3/2}\!\!
    \left(\frac{\lm}{1.3\U{mm}}\right)^{-1/2},
\end{equation}
and the equation $\gm\simeq4.0\left(m\rs{e}c^2/eB\lm\right)^{1/2}$
associates to each particle energy its typical synchrotron wavelength.

The ``transition region'' may be defined as the spectral region where the
diffusion time is comparable with the age of the Crab Nebula,
$\sim\!10^3\U{yr}$: the requirement of being located in between radio and mm
wavelengths gives a constraint on transverse size, typical field and diffusion
coefficient in filaments.
An observable effect of diffusion is the following:
if the particle density in these synchrotron filaments was originally higher
than in the surroundings, a transition to a diffusive case (spanning the
spectrum from long to short wavelengths) will level the particles density to
that of the interfilament medium and will therefore cause a fading of the
filament relative to the surrounding medium.

\subsection{Limits on the diffusion coefficient}

At the present stage any of the above mechanisms (either synchrotron burning or
transition to diffusion) can be invoked to account for the observed filament
dimming at mm wavelengths.
Anyway, even in the case of a transition in the diffusion regime, the field
inside filaments is required to be higher, in order to explain their
brightness.
We therefore suggest that synchrotron is the leading effect.
But further higher resolution mm observations are required to investigate the
physical conditions for individual filaments.

At any rate, the diffusion time of particles emitting at radio wavelengths
(say $\lm\sim5$~cm)
is required to be longer than the Crab Nebula age, which implies:
\begin{equation}
  \frac{\kp}{\kp\rs{B}}\la100\left(\frac{\tht\rs{F}}{1\arcsec}\right)^2
    \left(\frac{B}{0.3\U{mG}}\right)^{3/2}.
\end{equation}
Since typical transverse sizes of filaments are of the order of
$1\arcsec$, even for fields in filaments a few times higher than the average,
this upper limit on the diffusion coefficient for orthogonal diffusion in
individual filaments is inconsistent with the value ($\kp\sim\!10^4\kp\rs{B}$)
derived by comparing radial profiles at various wavelengths (Wilson
\cite{wil72a}, Amato et al.\ \cite{aea00}).

Imaging of individual filaments at mm wavelengths with arcsec resolution will
allow one to get more information on the origin of spectral differences between
filaments and interfilament medium, on the magnetic field in individual
filaments as well as on the effectiveness of particle diffusion.

\section{The nature of the component {\bf B}}

Free-free emission definitely cannot account for component {\bf B}.
In fact a very large mass ($\sim\!50 M_\odot$) of ionized matter is required to
give the measured mm excess flux.
In that case huge free-free emission should also appear in between infrared and
optical; moreover free-free absorption should become effective at wavelengths
above $\sim\!2$~m: but none of these effects has been observed.
Emission from a dust component is also unlikely (beyond the well known infrared
bump at a temperature of 46~K; Strom \& Greidanus \cite{sg92}).
In order to get the measured mm excess while maintaining the consistency with
far infrared photometric data, a dust temperature lower than 5~K is required,
together with a huge dust mass ($\sim\!100 M_\odot$, for standard dust).
These reasons, together with the fact that the map of the mm excess does not
match that of the thermal mass distribution in the Crab Nebula, lead us to
exclude a thermal nature for this emission.

Synchrotron emission is therefore the best candidate mechanism for component
{\bf B}, but also in this case very tight constraints on its spectrum must be
set from the data.
A ``non-thermal'' (i.e.\ negative) spectral index between radio and mm
wavelengths can be excluded; otherwise at radio wavelengths component {\bf B}
would affect considerably the flux, hence causing spatial inhomogeneities, in
the radio spectral index, above the quoted upper limits (Bietenholz et
al.\ \cite{bea97}).
The data are instead consistent with a low-energy cutoff in the energy
distribution of the emitting particles at a Lorentz factor
$\gm\rs{L}\sim\!1.5\E{4}$ (for $B=0.3\U{mG}$).

In the next section we shall examine morphological similarities between
component {\bf B}, as extracted from the mm map, and the region with an optical
spectrum flatter than the average.
If this similarity is attributed to the emergence of this component at optical
wavelengths and beyond, an emission spectrum with power law $\simeq-0.5$ is
required, implying an energy distribution above $\gm\rs{L}$ consistent with a
power law with index $-2$.
The comparison of such a spectrum with photometric data is presented in
Fig.~\ref{spfit}, with the warning that the parameters used for this model
spectrum are only indicative.
In this model component {\bf A} keeps the canonical $-$0.27 and $-$0.77
slopes in
radio and optical respectively (plus the known dust infrared bump): only the
position of the break has been slightly refined.

\begin{figure}
\caption{
A possible two-component spectral fit to photometric data from radio to optical
wavelengths.
Component {\bf A} plus the dust component are represented by the long-dashed
line, while the short-dashed line accounts for component {\bf B} (the solid line
showing the sum of both).
The vertical dot-dashed
line indicates the characteristic frequency associated with
the low-energy cutoff in the particles of component {\bf B}.
}
\label{spfit}
\end{figure}

The total number of emitting particles in component {\bf B} can be estimated as:
\begin{equation}
  N\rs{tot}\simeq2\E{48}\left(\frac{F\rs{low}}{100\U{Jy}}\right)
    \left(\frac{B}{0.3\U{mG}}\right)^{-1},
\end{equation}
where $F\rs{low}$ indicates its flux, measured near the low-energy cutoff.
This number is in nice agreement with what Kennel \& Coroniti (\cite{kc84b})
found, by fitting their wind model to the nebular spectrum between optical and
X rays: they give a rate of $2\E{38}\U{s^{-1}}$, while $N\rs{tot}$ can be
roughly estimated by multiplying this rate by the Crab age, $3\E{10}\U{s}$.

The most serious problem with otherwise successful wind models is their
inability to account for the large number of radio emitting particles.
The fact that the number of particles in component {\bf B} is instead in good
agreement with the Kennel \& Coroniti (\cite{kc84b}) prediction may indicate
that this mechanism is responsible for component {\bf B}, while a different
mechanism should be invoked for component {\bf A}.
Anyway a more detailed modelling, in the frame of wind models, of the
spectral-spatial properties of component {\bf B} is required in order to
clarify this point.

\section{Comparison with maps at other wavelengths}

The morphology of component {\bf B} (see also Fig.~\ref{compar}a) is
substantially different from that in the radio.
Except for a faint extended halo appearing in Fig.~\ref{compb}, but whose
existence is still uncertain, component {\bf B} is confined in a region with a
size of $\sim\!2\arcmin$.
It does not show the SE-NW elongation, typical of the radio map; instead, its
brightest feature is a narrow feature extending by about 1\arcmin\ in the NE-SW
direction, from whose centre an elongated feature protrudes towards NW, while
some further emission, with a broad shape, is located  on its SE side.

This morphology resembles that in X rays (shown in Fig.~\ref{compar}b with the
same resolution), or even better that of the region of the Crab Nebula
characterized by having flatter optical spectra (Fig.~\ref{compar}c shows the
spectral map in optical).
V\'eron-Cetty \& Woltjer (\cite{vw93}) have in fact already noticed a
similarity between the morphology of the optical spectral map and that of the
X-ray emission.

\begin{figure}
\caption{
Comparison of the mm map {\bf (a)} of component {\bf B} with maps at other
wavelengths (see the dashed reference line): both the X-ray map {\bf (b)} and
the optical spectral map {\bf (c)} show similar patterns; the radio map {\bf
(d)} seems instead anticorrelated with the main feature in component {\bf B}.
Each image is 256\arcsec\ in size.
}
\label{compar}
\end{figure}

What they have found can be qualitatively justified in terms of synchrotron
downgrading: if the particles are injected in the central regions, the X-ray
emission is confined to the inner part of the nebula because the emitting
particles have short synchrotron lifetimes; for the same reason, in the outer
part of the optical nebula synchrotron downgrading is responsible for the
spectral softening.

Component {\bf B}, as derived from the 1.3~mm map, now poses the following
riddle:  if the structure seen in X rays and in the optical spectral map is
uniquely determined by synchrotron downgrading, why a similar pattern can be
also seen (after subtracting the extrapolation from the radio) in the mm
spectral range, namely at wavelengths at which the emitting particles have
synchrotron lifetimes much longer than the age of the nebula?

Our conclusion is that synchrotron downgrading, which must anyway be present at
higher particle energies, in the Crab Nebula combines with the actual
coexistence of two different particle populations (which can be recognized as
component {\bf A} and component {\bf B}, whose spectra are shown in
Fig.~\ref{spfit}).
Imaging at different wavelengths in the mm range is required in order to
characterize more quantitatively the spectral properties of component {\bf B}.

From a comparison with the radio map (Fig.~\ref{compar}d, again reduced to the
same resolution of the mm map) it looks like the component {\bf B} and the
radio structure are anticorrelated, at least with reference to the main feature
in component {\bf B}.
Further observations with higher spatial resolution will allow a better
subtraction of the filaments, and a more detailed mapping of component {\bf
B}.
If they will confirm that component {\bf B} is actually located in a inner
region, there where radio emitting particles (i.e.\ component {\bf A}) present
instead a cavity, this result would indicate that the two components are also
spatially separated (with anyway a possible mixing in an intermediate region).
The presence of two spatially different components would simplify the modelling
of the dependence with wavelength of the Crab Nebula size; on the other hand it
remains to be explained how the long-lived particles in component {\bf B} can
be effectively confined in the inner part of the Crab Nebula.

\section{Effects of time variability}

When discussing whether the detection of component {\bf B} could be a mere
effect of an observational bias, we have limited our analysis to instrumental
effects.
However the comparison for the 1.3~mm with a radio map obtained at a different
epoch is not justified, if time variations of the source could account for the
observed differences.

In fact short-time variability has been observed, in the Crab Nebula, in
optical about the wisps region (Hester et al.\ \cite{hea95}) and in X rays
about the torus (Greiveldinger \& Aschenbach \cite{ga99}).
Moreover, some small-scale wisp-like features, for which a peculiar radio
spectral index had been previously quoted (Bietenholz \& Kronberg \cite{bk92}),
have been actually discovered to vary (Bietenholz et al.\ \cite{bea01}) with
timescales of the order of one yr.

One may then wonder whether time variability may account for the changes in
morphology between the radio and mm we have used, and that we have interpreted
instead as due to inhomogeneities in the spectral index.

We can exclude that what we have seen is a mere effect of time variability,
for various reasons.
The radio variability pattern, as shown by Bietenholz et al.\ (\cite{bea01}),
presents a wavy structure with near arcsec scalelength: when smeared to
10\arcsec\ resolution (as for our mm map), the time varying structures are very
likely to be averaged out completely.
Moreover, our component {\bf B} is spatially more extended than the time
variable region shown by Bietenholz et al.\ (\cite{bea01}), which does not
extend further than the NE-SW elongated feature in the innermost part of
component {\bf B} (see Fig.~\ref{compb}): even not considering its outer halo,
the two extensions in the SE and NW sides do not present any noticeable radio
variability.

An overall flux variability involving the inner $\sim\!1\farcm6$ of the nebula is
also unlikely because such a large change would affect with ripples the secular
flux evolution of the Crab Nebula, but no effect of this kind has been detected
yet (Aller \& Reynolds \cite{ar85}).
On the other hand it should be clear that synchrotron lifetimes of mm emitting
particles are long, and therefore, if a detectable short-time flux increase
were produced by a burst in the particle injection, this burst must inject a
number of particles comparable with all particles injected in the past history
of the Crab Nebula; moreover the effects of one such burst will leave for a
very long time.
Therefore we exclude as very remote the eventuality that time evolution be
responsible of what we have interpreted as real spectral inhomogeneities.

Bietenholz et al.\ (\cite{bea01}) claim that the radio detection of dynamical
structures in the same region where optical and X-ray wisps have been detected
suggests a common acceleration mechanism.
In a strict sense what they found is that at least ``some'' radio emission has
the same origin as the X-ray emission: this result is consistent with ours,
since even in the case of a low-energy cutoff in the particle distribution
component {\bf B} will emit in radio, following the low-frequency $\nu^{1/3}$
synchrotron law.

At radio wavelengths it will be crucial to map the spectral index of the inner
regions by using maps obtained at nearly the same time: only in this way it
will be possible to tell whether all spectral inhomogeneities accounted for by
Bietenholz \& Kronberg (\cite{bk92}) are an effect of time variability, or
whether the flatter spectra they found coincident to some radio wisps will be
confirmed.

\section{Conclusions}

We have measured inhomogeneities in the spectral index between radio and mm
wavelengths, and we have shown that they could be better explained in terms of
the emergence of a further synchrotron component, undetected at radio
wavelengths, which is located in the inner part of the nebula.

The transition in size and shape of the Crab Nebula, moving from the radio to
the X-ray spectral range, is qualitatively explained in terms of synchrotron
downgrading.
But the fact that size and shape of component {\bf B} at mm wavelengths
resemble those seen in X rays cannot be explained in that way, since mm
emitting particles are subject to only minor synchrotron losses.
We suggest instead that two different synchrotron components coexist in the
Crab Nebula and that the morphological transition taking place from radio to X
rays requires a change of the relative importance of the two components.

In order to fit the data, the energy distribution of particles emitting in
component {\bf B} requires to have a low-energy cutoff.
With this, the total number of particles in component {\bf B} are in agreement
with what predicted by Kennel \& Coroniti (\cite{kc84b}) model.

Finally, in synchrotron filaments we have found the evidence for a spectral
break at a frequency lower than that averaged over the whole nebula.
Although there may be some effects related to particle diffusion through
filaments, we take a magnetic field in filaments higher than in their
surroundings as the main cause of this effect.

\begin{acknowledgements}
We wish to thank E.~Amato, J.~Arons, J.~Hester, L.~Woltjer for interesting
discussions.
We are grateful to M.~Bietenholz for having provided us with a high quality
radio map of the Crab Nebula.
This work has been partly supported by the Italian Ministry for University and
Research under Grant Cofin2001--02--10, and by the National Science Foundation
under Grant No.~PHY99--07949.
\end{acknowledgements}

\appendix

\section{procedure used to separate the filament and amorphous components}

\begin{figure}
\caption{
The filament ({\bf f}; top panel) and amorphous ({\bf a}; bottom panel)
components, at full resolution, extracted from the radio map using the
procedure described in the Appendix.
The two maps are displayed using the same intensity levels: the grey scales,
shown on the right, range from 0 to $15\U{mJy\,arcsec^{-2}}$, in steps of
$1\U{mJy\,arcsec^{-2}}$.
The {\bf f} and {\bf a} components account respectively for 30\% and 70\% of
the total flux at 1~GHz.
The map of filaments has been then downgraded to the IRAM 30-m telescope
resolution (Fig.~\ref{fila}).
}
\label{twocomp}
\end{figure}

In this appendix we describe in detail the procedure we have followed to
extract from the the high-resolution map at 20~cm both the structure of the
filaments network (component {\bf f}) and that of the amorphous part (component
{\bf a}).
This separation is based only on morphological criteria and relies on the fact
that the typical scale lengths of the two components are very different.
As already stated in the text, our morphological requirements are: component
{\bf a} must look very smooth; component {\bf f} must instead contain mostly
high spatial frequencies; negative peaks must appear neither in the map of
component {\bf f} nor in that of {\bf a}; the total flux of the filaments
component must be as high as possible.

A standard method to separate components characterized by different spatial
frequencies is linear filtering.
However this method introduces negative sidelobes, hence giving a null
integrated flux for the filaments.
For the extraction of the filaments contribution we have then applied a
non-linear iterative procedure, with the goal of gradually ``cleaning'' the
component {\bf a} from {\bf f}.
We start with {\bf a$_0=M$}, the original map.
In each step we convolve {\bf a$_{i-1}$} by a 2-D gaussian with a given width
$\sg\rs{g}$, and subtract the resulting image from $M$, using a gain $G_i$; we
then clip negative peaks to obtain {\bf f$_i$} and evaluate {\bf a$_i$} as
$M-${\bf f$_i$}.
At the final step $G_i=1$ is required, in order to subtract completely the
filaments.
We have obtained good results (Fig.~\ref{twocomp}) with just 2 steps, using
$\sg\rs{g}=20\arcsec$, $G_1=0.75$ and $G_2=1$.
This procedure has been applied to the full resolution radio map, and only
afterwards the filaments map has been convolved to the resolution of our 1.3~mm
map (the final map is shown in Fig.~\ref{fila}).


\end{document}